\documentclass[conference]{IEEEtran}
\IEEEoverridecommandlockouts
\usepackage{cite}
\usepackage{amsmath,amssymb,amsfonts}
\usepackage{algorithmic}
\usepackage{graphicx}
\usepackage{textcomp}
\usepackage{xcolor}
\usepackage{tabularx}
\usepackage{listings}
\usepackage{nopageno}

\def\BibTeX{{\rm B\kern-.05em{\sc i\kern-.025em b}\kern-.08em
    T\kern-.1667em\lower.7ex\hbox{E}\kern-.125emX}}

\begin{document}

\thispagestyle{empty}

\begin{huge}
IEEE Copyright Notice
\end{huge}

\vspace{5mm} 

\begin{large}
Copyright ©2020 IEEE
\end{large}

\vspace{5mm} 

\begin{large}
Personal use of this material is permitted. Permission from IEEE must be obtained for all other uses, in any current or future media, including reprinting/republishing this material for advertising or promotional purposes, creating new collective works, for resale or redistribution to servers or lists, or reuse of any copyrighted component of this work in other works.
\end{large}

\vspace{5mm} 

\begin{large}
\textbf{Accepted to be published in:} 2020 3rd International Conference on Information and Computer Technologies (ICICT) http://www.icict.org/
\end{large}

\vspace{5mm} 

\begin{large}
DOI 10.1109/ICICT50521.2020.00068
\end{large}

\newcolumntype{L}[1]{>{\raggedright\arraybackslash}p{#1}}
\newcolumntype{C}[1]{>{\centering\arraybackslash}p{#1}}
\newcolumntype{R}[1]{>{\raggedleft\arraybackslash}p{#1}}

\title{Is Cryptojacking Dead after Coinhive Shutdown?}

\makeatletter
\newcommand{\linebreakand}{%
  \end{@IEEEauthorhalign}
  \hfill\mbox{}\par
  \mbox{}\hfill\begin{@IEEEauthorhalign}
}
\makeatother
\author{\IEEEauthorblockN{Said Varlioglu}
\IEEEauthorblockA{\textit{School of Information Technology} \\
\textit{University of Cincinnati}\\
Cincinnati, Ohio, USA \\
varlioms@mail.uc.edu}
\and
\IEEEauthorblockN{Bilal Gonen}
\IEEEauthorblockA{\textit{School of Information Technology} \\
\textit{University of Cincinnati}\\
Cincinnati, Ohio, USA \\
bilal.gonen@uc.edu}
\linebreakand 
\IEEEauthorblockN{Murat Ozer}
\IEEEauthorblockA{\textit{School of Information Technology} \\
\textit{University of Cincinnati}\\
Cincinnati, Ohio, USA \\
ozermm@ucmail.uc.edu}
\and
\IEEEauthorblockN{Mehmet F. Bastug}
\IEEEauthorblockA{\textit{Department of Interdisciplinary Studies} \\
\textit{Lakehead University}\\
Orillia, Ontario, CA \\
mbastug@lakeheadu.ca}

}

\maketitle

\begin{abstract}
Cryptojacking is the exploitation of victims' computer resources to mine for cryptocurrency using malicious scripts. It has become popular after 2017 when attackers started to exploit legal mining scripts, especially Coinhive scripts. Coinhive was actually a legal mining service that provided scripts and servers for in-browser mining activities. Nevertheless, over 10 million web users had been victims every month before the Coinhive shutdown that happened in Mar 2019. This paper explores the new era of the cryptojacking world after Coinhive discontinued its service. We aimed to see whether and how attackers continue cryptojacking, generate new malicious scripts, and developed new methods. We used a capable cryptojacking detector named CMTracker that proposed by Hong  et  al. in 2018. We automatically and manually examined 2770 websites that had been detected by CMTracker before the Coinhive shutdown. The results revealed that 99\% of sites no longer continue cryptojacking. 1\% of websites still run 8 unique mining scripts. By tracking these mining scripts, we detected 632 unique cryptojacking websites. Moreover, open-source investigations (OSINT) demonstrated that attackers still use the same methods. Therefore, we listed the typical patterns of cryptojacking. We concluded that cryptojacking is not dead after the Coinhive shutdown. It is still alive, but not as attractive as it used to be. \footnote{Preprint.  This paper was presented in 2020 3rd International Conference on Information and Computer Technologies (ICICT); http://www.icict.org/downloads/PROGRAM.pdf} 

\end{abstract}


\section{Introduction}
There are two ways to gain a cryptocurrency. The first one is to buy it and the second one is to mine it. Successful miners obtain new cryptocurrency as a reward through the mining process. The mining is to solve complicated mathematical problems that require processing a large number of hash-like computation workload \cite{Hong2018} \cite{Eskandari2018}. However, mining activity is a tough process for a regular computer user due to its financial and technical burdens. 
The mining process requires high CPU usage, thus heavy electricity consumption and receiving high bills. 
As the interest in cryptocurrencies has surged, attackers have started to use hidden malicious scripts to mine cryptocurrencies without any expense.
This attack is named ``Cryptojacking'' consisting of two words; Cryptocurrency and Hijacking.

There are two types of cryptojacking attacks. 
First, attackers put malicious hidden scripts on their websites. When a user visits thesehis website, the malicious script is loaded to the victim user's computer and exploits the victim's computing power. 
Second, an attacker injects a script to a victim's website. 
In this second type attack, an attacker gain benefit using a victim's injected website which exploits users' computers who visit this website.

Cryptojacking attacks emerged and became vastly popular after 2017.
Symantec reported that they blocked approximately 16 million attacks in 2017 and blocked 69 million attacks in 2018 with more than four times increase \cite{Dwight2019}. As an example of this crime effect, Japan police arrested 16 individuals suspected of exploiting victim's computer to mine for cryptocurrency in 2018 \cite{Sudo2018}.

A victim may suffer from cryptojacking concerning not only high electricity bills but also computer's overall performance. A study demonstrated that a cryptojacking website affects a victim’s computer by increasing temperature up to 52.8\%, decreasing performance up to 57\% and multiplying CPU usage up to 1.7x \cite{Papadopoulos2019}.

Coinhive was a company that used to provide a script-code to enable website owners mine Monero cryptocurrency \cite{monero} holding 30\% of all mined XMR for itself \cite{Eskandari2018}. Since Monero claims that it is an untraceable cryptocurrency, it became popular with the ``easy to use'' feature of Coinhive scripts after 2017.

Nevertheless, attackers immediately started to exploit those scripts to mine cryptocurrencies. On Google, ``Monero'', ``Coinhive'' and ``Cryptojacking'' words search frequencies significantly increased. Even though Coinhive started to provide this service with a benign purpose, the script was maliciously used by thousands of attackers or injected into victims' websites within a couple of months \cite{Musch2018}. 
Studies suggested that 81\% of the cryptojacking websites used scripts provided by Coinhive \cite{Saad2018}.
Symantec reported that ``statdynamic.com/lib/crypta.js'' script, which connects Coinhive servers, was found even on Microsoft Store \cite{Guo2019}.

Before Coinhive shutdown the service, Hong et al. \cite{Hong2018} developed a novel detection system named CMTracker. Its detection relies on behaviour-based profilers consisting of hash-based and stack-based profilings. It monitors each webpage for less than three seconds using two runtime profilers for automatically tracking malicious scripts and related domains. According to CMTracker, there were 868 Cryptojacking websites in Alexa top 100K list. The discovery was over 260\% more comparing with the latest public reports. Also, they found 2,770 malicious websites from 853,936 popular web pages.

On March 8, 2019, Coinhive stopped its service claiming the economic reasons regarding to new Monero hard fork impacted their profitability. 
Also, Monero XMR value has fallen by as much as 85 percent over the past year \cite{Cimpanu2019}. After those issues, the company removed even the website ``Coinhive.com'' on Apr 29, 2018, by leaving this message: \textit{``Some of you will be surprised. The decision has been made. We will discontinue our service on March 8, 2019. It has been a blast working on this project over the past 18 months, but to be completely honest, it isn't economically viable anymore.''} \cite{CoinhiveTeam2019}.

Coinhive made a decision affecting the cryptojacking world significantly. In this paper, we are exploring the consequences of this decision. We will draw a picture of this new era in cryptojacking. We aimed to see that, after eight months of Coinhive shutdown, whether the attackers continue this type of cyberattacks, generate new malicious scripts, use new methods or not.

We used two methods for examination. First, we used a capable Cryptojacking detector named CMTracker that has been developed by Hong et al. \cite{Hong2018}. Second, we manually examine 2,770 websites that had been detected before by CMTracker to make sure whether CMTracker fails or not on identifying cryptojacking websites and malicious scripts after Coinhive. 

The results suggested that 99\% of websites that had been detected by CMTracker before, no longer continue cryptojacking. 1\% of websites still continue mining with current scripts. In addition, examining those websites manually shows us that CMTracker can still detect the cryptojacking websites. That means attackers still use the same methods. Since the attackers still use the same methods, we observed that cryptojacking patterns are clearly noticeable. We listed the patterns of cryptojacking in the examination section.

Although the majority of cryptojacking websites (99\%) stopped their activities, we were able to track 8 unique mining scripts derived from the remaining 1\% cryptojacking websites.
As a result, we detected 632 unique cryptojacking websites. Some of them have millions of visitors per year. In conclusion, we would not say that cryptojacking is dead after Coinhive shutdown. It is still alive but not attractive for the present.

\section{Examination of Cryptojacking Websites}
We examined 2,770 websites located in CMTracker GitHub address. The results demonstrated that, after Coinhive, while 1\% of attackers continue their activities, 99\% percent of them has stopped. Also, 300 websites out of 2,770 websites still try to connect ``statdynamic.com/lib/crypta.js'' script located in dead Coinhive servers currently.

CMTracker generated several JSON files for each page. It detected total of 155 webpages. At first, hash-based profiler detected 114, stack-based profiler detected 41 cryptojacking webpages. However, when we examine those webpages, we noticed that there are duplications in the results. Once we clear the duplications, we observed that while the hash-based profiler found 14 unique web pages, the stack-based profiler found 10 unique pages. Finally, we concluded that 24 webpages out of 2,770 websites still execute Cryptojacking mining scripts in the background without users' consent as seen in Table \ref{table1}. 

We also  examined  the  24  websites to find the unique mining scripts. As seen in Table \ref{table2}, we revealed 8 running  live  Cryptojacking scripts.

\begin{table}[h]
  \caption{Results of CMTracker after Coin-Hive Shutdown}
  \centering
  \label{table1}
  \setlength\tabcolsep{1mm}
  \begin{tabular}{|p{20mm}|p{18mm}|C{14mm}|C{14mm}|C{8mm}|}
    \hline
    &Domain List&Hash-based Detection&Stack-based Detection&Total\\
    \hline
    Raw Results&2,770 Websites&114&41&155\\
    \hline
    Cleaned Results&2,770 Websites&14&10&24\\
    \hline
  \end{tabular}
\end{table}

Moreover, the results suggested that the hash-based profiler is more effective than stack-based profiler. In 4 separate tests on 24 websites, the hash-based profiler was able to detect all of websites, however, the stack-based profiler detected only 10 websites.

\begin{table}[h]
    \centering
  \caption{Mining Scripts Derived From 24 Websites}
  \label{table2}
  \setlength\tabcolsep{1mm}
  \begin{tabular}{|p{6mm}|p{60mm}|}
    \hline
    \textbf{N}&\textbf{Mining Script}\\
    \hline
    1&bitcoin-pay.eu/perfekt/perfekta.js\\
    \hline
    2&easyhash.de/tkefrep/tkefrep.js\\
    \hline
    3&enaure.co/javas.js\\
    \hline
    4&hashing.win/46B8.js\\
    \hline
    5&lasimakiz.xyz/sadig6.js\\
    \hline
    6&minero.cc/lib/minero.min.js\\
    \hline
    7&uvuvwe.bid/jo/jo/miner\_compressed/webmr.js\\
    \hline
    8&webminepool.com/lib/base.js\\
    \hline
  \end{tabular}
\end{table}

Through a manual survey of 500 websites out of the 2,770 cryptojacking webpages detected by CMTracker, 344 were found to be no longer running mining scripts, 92 were shut down, 58 were still attempting to connect Coinhive using ``coinhive.min.js'' script, and 6 were using other mining scripts. Out of 500 websites we surveyed manually, 68.8\% of websites removed all hidden malicious scripts, 18.4\% of websites have stopped their service, 11.6\% of websites still use Coinhive scripts, 1.2\% of websites still continue Cryptojacking activities as seen in Table \ref{table3}.

\begin{table}[h]
    \centering
  \caption{Distribution of Cryptojacking Activity on 500 Websites}
  \label{table3}
  \setlength\tabcolsep{1mm}
  \begin{tabular}{|L{45mm}|R{15mm}|}
    \hline
    \textbf{Observation}&\textbf{Percentage}\\
    \hline
    No Malicious Hidden Script&68.8\%\\
    \hline
    No Web Service&18.4\%\\
    \hline
        Still Contain Coinhive Scripts&11.6\%\\
    \hline
        Running Other Live Malicious Scripts&1.2\%\\
    \hline
  \end{tabular}
\end{table}

After the Coinhive shutdown, revisiting and examination of the original 2,770 websites demonstrated that malicious cryptojacking activities have significantly declined by 99\%. Only 24 websites out of 2,770 websites currently contain running mining scripts.

\section{OSINT on Detected Cryptojacking Scripts}
By examining the 24 websites that run live Cryptojacking scripts, we revealed 8 unique mining scripts. We conducted OSINT searching and tracking down those scripts on PublicWWW.com that is the source search engine. The result suggested that 632 cyptojacking websites contain those Cryptojacking mining scripts as seen in Table \ref{table4}. 

On the other hand, according to "PublicWWW.com" records, there are over 7,000 websites still contain the Coinhive mining scripts as of Sep 2019. However, it was over 30,000 as of Nov 2017 \cite{Mursch2017}. Thus, over 23,000 websites have erased the Coinhive scripts after the service stopped.

\begin{table}[h]
  \caption{Number of websites using the detected scripts for mining activity}
  \label{table4}
  \setlength\tabcolsep{1mm}
  \begin{tabular}{|L{54mm}|R{26mm}|}
    \hline
    \textbf{Mining Scripts}&\textbf{Number of websites}\\
    \hline
    minero.cc/lib/minero.min.js&275\\
    webminepool.com/lib/base.js&121\\
    hashing.win/46B8.js&96\\
    */perfekt/perfekt.js&85\\
    */tkefrep/tkefrep.js&30\\
    enaure.co/javas.js&17\\
    lasimakiz.xyz/sadig6.js&4\\
    uvuvwe.bid/jo/jo/miner\_compressed/webmr.js&4\\
    \hline
    \textbf{Total}&\textbf{632}\\
    \hline
  \end{tabular}
\end{table}

In addition, we observed that a user in a cybersecurity blog complained about the mining script injection attack: \textit{``I realize someone is putting crypto miner on my pages. If I delete it several days later it returns. My question is, how is it possible for an attacker to inject code directly into the document? Wouldn't you require server credentials to access and edit files on the server?''.} \cite{stackexchange2018} ``perfekt.js'' mining script was used for this purpose. This incident demonstrates that attackers uses same mining scripts that have been produced another attacker.

\section{Patterns of Cryptojacking}
Since Coinhive was originally created to serve as a browser-based cryptominer and it was not written with obfuscation in mind, it was easy to detect even when it runs without the visitor's knowledge. On the other hand, other scripts developed after Coinhive use code obfuscation to avoid detection. Nevertheless, even if an attacker use obfuscation techniques to avoid detection systems, a regular user can easily detect a cryptojacking website based on common patterns in a website.

WebSockets, WebWorkers and WebAssembly (wasm) connections underlie a Cryptojacking activity to get the connections robust \cite{Neumann2018, Musch2018, Bijmans2019}. The existence of such connections may indicate a cryptojacking activity. Miners use WebSocket (wss://) protocol to make the user keep connected. WebSocket is used to establish a connection with the server. It is a different protocol from HTTP. It provides full-duplex communication channels over a single TCP connection. Thus, it is used as a continuous/persistent connection between a user and a server.

Besides that, a cryptojacking website runs four JavaScript workers as threads connecting blob links. Four JavaScript workers redirect the user to miner deployers. A blob link is seen as four different links with octet stream type in Chrome network activities. Thus, four java workers are employed by a direct server (mining pool) or another website (miner deployer). Also, those call stack threads are called ``Long Tasks'' by Chrome. 

Cryptojacking websites redirect the user mining deployer or mining pool using malicious scripts. If a script code is not obfuscated, its code usually contain ``var miner'', ``throttle'', ``start miner'' phrases. ``Throttle'' function is used to limit the maximum CPU usage during mining to avoid high CPU usage detection.

Hardware Performance Counters (HPCs) are used to detect a Cryptojacking activity on a browser. 
Since crypto-mining worsen hardware performance dramatically by multiplying CPU usage, increasing temperature, HPCs are another indicator of attacks.

According to a recent research, Cryptojacking becomes more profitable than ads only when a user remain in a  website longer than 5.53 minutes \cite{Papadopoulos2019}. We observed that Cryptojackers mostly use free movie websites to try to keep the victims connected the website for a long time.

Finally, we can list those patterns in six categories:
\begin{enumerate}
    \item WebSockets, WebAssembly connections
    \item Four WebWorkers (JavaScript workers)
    \item Long Tasks in Call Stack Threads
    \item The existence of ``var miner'', ``throttle'', ``start miner'' phrases in the redirection links
    \item CPU and battery usage (hardware performance counters)
    \item Free movie contents
\end{enumerate}{}

\section{Case Studies}
After Coinhive closure, in order to figure out other Cryptojacking websites behavior in this new era, we examined the websites below. 

``cinecalidad.to'' is a free movie websites which has millions of visitors per year. This website employs a cryptojacking script by connection mining deployer: ``enaure.co/javas.js''. Enaure.co is a domain that hosts mining scripts for cryptojacking. ``Cinecalidad.to'' connects ``Enaure.co'', and ``Enaure.co'' connects ``Moneroocean.stream'' mining pool website that assign mining tasks and return rewards to miners. The network traffic of mining in ``cinecalidad.to'' relies on a WebSocket and WebAssembly connection to get the victim permanently connected to the server. It performs ``long task'' with four JavaScript workers. JavaScript workers run on four blob links. Also, Google chrome developer tools alert the user under Network – Timing section: \textit{``Request is not finished yet''} which means that there is a continuous connection between victim and attacker mining server. The main mining connection methods contain the original website, the mining deployer website, the mining pool website and continuous connection Wasm and WebSocket protocols as seen in Table \ref{table5}. In addition to that, mining scripts employ four javascript workers as seen in Table \ref{table6}.

\begin{table}[h]
    \centering
  \caption{Network Traffic of Mining in ``cinecalidad.to''}
  \label{table5}
  \setlength\tabcolsep{1mm}
  \begin{tabular}{|L{70mm}|}
    \hline
    Origin: https://www.cinecalidad.to\\
    Request URL: wss://www.enaure.co:8181/\\
    Identifier: ``handshake'', pool: ``moneroocean.stream''\\
    pool: ``moneroocean.stream''\\
    Sec-WebSocket-Key: ``6CzfNw...''\\
   Wasm://wasm/wasm-0007c7f6/wasm\\
    \hline
  \end{tabular}
\end{table}

Furthermore, the script codes includes the common cryptojacking phrases such as ``startMining'', ``throttleMiner'',``stopMining'', ``wasmSupported'', ``pool'', ``addWorkers'', ``deleteAllWorkers'', ``new Blob''. In Table \ref{table7}, we listed a regular computer hardware reaction while connecting to ``Cinecalidad.to''. According to Alexa, the rank of cinecalidad.to is 597 as of Oct 10, 2019. Millions of people visit it per year. The users are from Mexico 22.7\%, Venezuela 18.0\% and Argentina 9.4\% , etc. \cite{Musch2018}.

\begin{table}[h]
  \caption{Four Blob Mining Workers}
  \label{table6}
  \setlength\tabcolsep{1mm}
  \begin{tabular}{|L{85mm}|}
    \hline
blob:https://www.cinecalidad.to/f0ce0413-255c-4cbb-92ae-d0c5da05f910\\
    \hline
blob:https://www.cinecalidad.to/5b1f270c-035e-441f-aaf7-b7c337e5db25\\
    \hline
blob:https://www.cinecalidad.to/34b066c4-dffa-4575-93f0-c2e49fb1ed44\\
    \hline
blob:https://www.cinecalidad.to/aebe1ff1-17c8-48db-8fdb-d725d7530be5\\
    \hline
  \end{tabular}
\end{table}

\begin{table}[h]
  \caption{Computer Hardware Reaction During Cryptojacking}
  \label{table7}
  \setlength\tabcolsep{1mm}
  \begin{tabular}{|p{17mm}|p{13mm}|p{16mm}|C{13mm}|C{14mm}|}
    \hline
    Cryptojacking&Speed&Core Speed&Multiplier&Temperature\\
    \hline
    No&0.87 GHz&798.7 MHz&x8.0&47 °C\\
    \hline
    Yes&2.60 GHz&2595.7 MHz&x26.0&85 °C\\
    \hline
  \end{tabular}
\end{table}

The other website that we examined is ``piroozgar.com''. It is a shopping website with a 169,245 alexa global rank. This website employs a cryptojacking script named ``webminepool.com/lib/base.js''. It uses a different method from ``cinecalidad.to'' that we examine above. ``piroozgar.com'' gets the user connected the mining pool directly. Thus, mining pool performs already a mining deployer operation by assigning mining tasks to the user. However other operations are same. The network traffic of mining in ``piroozgar.com'' relies on a WebSocket and WebAssembly connection.  It also runs \textit{``long task''} with four JavaScript workers with blob functions. Google chrome developer tools alert the user under console: \textit{``Request is not finished yet'' and ``Violation: Forced reflow while executing JavaScript took 68ms, 72ms and 87ms''}. We demonstrate its a call stack operation on one of four java workers in Figure \ref{fig1}. It has also cryptojacking phrases such as ``new Miner'', ``new Miner'', ``throttleMiner''.

\begin{figure}
    \centering
    \includegraphics[width=1\columnwidth]{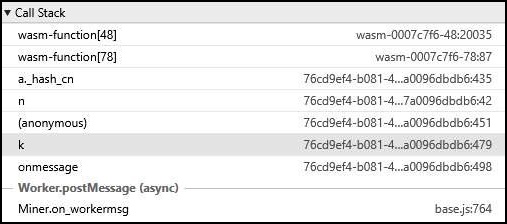}
    \caption{Sample Call Stack Operation of a Cryptojacking Website}
    \label{fig1}
\end{figure}

Before Coinhive, Hong et al. \cite{Hong2018} also conducted real-world examples as case studies. In order to see what they act after Coinhive shutdown, we examined them again. First one is an automobile website, ``planete-auto-entrepreneur.com'' is still existing. Second one ``dlight.ir'' which used both Coinhive and Crypto-Loot scripts is still existing and try to connect coinhive and coin-have mining servers. Although all of them had also different mining scripts from Coinhive, interestingly, they no longer run mining scripts. The others, ``planetatvonlinehd.com/dark-temporada-1'' and ``bookstore.investmentu.com'' which is not working. 

The other interesting and popular example that we examined is ``thepiratebay.org''. This website provides free digital content of entertainment media and software. It used to run mining scripts during Coinhive is alive with a small note that alerts the users about cryptomining in the bottom of main page. The examinations were conducted on Mar 10, 2019 and on Sep 16, 2019. In the first examination, it used to run a mining script. However in the second examination, thepiratebay does not use mining scripts anymore.

\section{Related Works}
Researchers have proposed novel detection systems in the literature using the common patterns of cryptojacking that we listed above. Tahir et al. \cite{Tahir2019} proposed a machine learning approach based on hardware-assisted profiling of browser code in real-time.
Rodriguez and Posegga \cite{ParraRodriguez2018} developed several approaches to detect in-browser mining  monitoring dynamically API calls and resource consumption.
Hong et al. \cite{Hong2018} proposed a behavior-based detector with two runtime profilers for automatically tracking Cryptocurrency Mining scripts and their related domains.
Ko et al. \cite{Ko2018} developed a cryptojacking detection solution using a dynamic analysis-based that uses a headless browser.
Wang et al. \cite{Wang2018a} proposed a detection system named SEISMIC (Secure In-lined Script Monitors for Interrupting Cryptojacks) based on monitoring Wasm in-line scripts \cite{Wang2018a}.

In addition, Bijmans et al. \cite{Bijmans2019}, Eskandar \cite{Eskandari2018}, Marchetto and Liu \cite{Marchetto2019}, Saad et al. \cite{Saad2018}, Musch et al. \cite{Musch2018} examined in-browser mining of cryptocurrencies in all its aspects. They revealed its deployment, expansion, trends, organized structures and features statically and dynamically.

\section{Results and Conclusion}
The results demonstrated that 99\% of websites that had been detected by CMTracker before, they no longer use hidden malicious scripts. Only 1\% of websites still run cryptominer scripts. 24 webpages out of 2,770 websites have those scripts. However, when we track 8 unique mining scripts derived from those 1\% webpages, we detected 632 other Cryptojacking websites. A detail examination shows some specific patterns of Cryptojacking and we listed them in 6 categories. Thus, we would say that, it is not hard to detect a Cryptojacking website. Finally, the existence of 632 cryptojacking websites give us enough evidence to say that Cryptojacking did not end after Coinhive shutdown. It is still alive but not as appealing as it was before. It became less attractive not only because Coinhive discontinued their service, but also because it became a less lucrative source of income for website owners. For most of the websites, ads are still more profitable than mining. Diminishing value of Monera and stricter monitoring by advertisements networks \cite{Bijmans2019} also discouraged website owners to rely on cryptojacking. We do not know how many new scripts have been created after Coinhive, and how many websites use them intentionally or unintentionally. Apparently, a new extensive research is needed to examine the new era of Cryptojacking.

\section{Acknowledgments}
We want to thank Ian Keller, who helped us a lot by fixing and deploying the python codes of the examination.

\bibliographystyle{IEEEtran}
\bibliography{references}

\end{document}